\documentclass[12pt]{article}

  \usepackage{graphicx}
  \usepackage{epsfig}
  \usepackage{amssymb}
  \usepackage{subfigure}

\evensidemargin - 1.truecm
\begin{document}

\newcommand{\be}{\begin{equation}}
\newcommand{\ee}{\end{equation}}
\newcommand{\nn}{\nonumber}
\newcommand{\bea}{\begin{eqnarray}}
\newcommand{\eea}{\end{eqnarray}}
\newcommand{\bfig}{\begin{figure}}
\newcommand{\efig}{\end{figure}}
\newcommand{\bc}{\begin{center}}
\newcommand{\ec}{\end{center}}
\newcommand{\bd}{\begin{displaymath}}
\newcommand{\ed}{\end{displaymath}}

\begin{titlepage}
\nopagebreak
%
\vspace*{-1.5cm}                        
\vskip 3.5cm
\begin{center}
\boldmath
{\Large \bf Heavy-Quark Form Factors \\
\vspace*{3mm}
and Threshold Cross Section
at ${\mathcal O}(\alpha_{S}^{2})$}\unboldmath
\vskip 1.5cm
{\large  W.~Bernreuther$\rm \, ^{a \,}$
}
{\large  R.~Bonciani$\rm \, ^{b, \,}$\footnote{Speaker. Email: 
{\tt Roberto.Bonciani@ific.uv.es}}}
{\large T.~Gehrmann$\rm \, ^{c \,}$
},
{\large R.~Heinesch$\rm \, ^{a \,}$
}, \\[2mm] 
{\large T.~Leineweber$\rm \, ^{a \,}$
}, 
{\large P.~Mastrolia$\rm \, ^{d \,}$
}, and
{\large E.~Remiddi$\rm \, ^{f \,}$
}
\vskip .7cm
{\it $\rm ^a$ Institut f\"ur Theoretische Physik, RWTH Aachen,
D-52056 Aachen, Germany} 
\vskip .3cm
{\it $\rm ^b$ 
Dep. de F\'{\i}sica Te\`orica, IFIC --
CSIC, Univ. de Val\`encia,
E-46071 Valencia, Spain} 
\vskip .3cm
{\it $\rm ^c$ Institut f\"ur Theoretische Physik, 
Univ. Z\"urich, CH-8057 Z\"urich, Switzerland}
\vskip .3cm
{\it $\rm ^d$ Dep. of Physics and Astronomy, UCLA,
Los Angeles, CA 90095-1547} 
\vskip .3cm
{\it $\rm ^f$ Dip. di Fisica dell'Universit\`a and INFN Sez. di Bologna, 
I-40126 Bologna, Italy} 
\end{center}
\vskip .4cm

\begin{abstract}

During the last year, analytic expressions for the two-loop QCD 
corrections to the form factors for the vector, axial-vector, scalar 
and pseudo-scalar vertices involving a pair of heavy quarks, $Q \bar{Q}$, 
were calculated. The results are valid for arbitrary momentum transfer and 
mass of the heavy quarks. These form factors have a number of applications, 
including anomalous couplings, the $e^{+}e^{-} \to Q \bar Q$ cross section, 
and the forward-backward asymmetry of heavy quarks. Here the $Q {\bar Q}$ 
threshold cross section is presented with some new second order axial vector 
contributions.

\flushright{
        \begin{minipage}{12.3cm}
\hspace*{18.5mm} Heavy quarks.
        \end{minipage}        }
\end{abstract}

\vskip 1.cm

\flushleft{\it HEP2005 International Europhysics 
           Conference on High Energy Physics\\
           July 21st - 27th 2005 \\
           Lisboa, Portugal}
\vfill
\end{titlepage}

In the next years particle physics will receive a big boost, due mainly 
to the start of activity of the Large Hadron Collider (LHC) that will 
explore physics at the TeV scale. One of the purposes of such a program 
is the understanding of the electroweak symmetry breaking and in particular
a possible confirmation of the Higgs mechanism, via the discovery of the
Higgs boson, the last still missing particle of the Standard Model 
(SM). In this context, an important role will be played by heavy quarks. 
This is the sector of the SM where a possible deviation from the Higgs 
mechanism could be detected first (in particular for the top quark that 
has a large mass and thus a large coupling to the Higgs). 
Therefore, a precise theoretical determination of observables concerning 
the $b$- and $t$-quark production and decay processes is mandatory in order 
to match the precision that will be required by the Tevatron, LHC and by
an International Linear Collider.

A step in this direction was made in the last year with the calculation 
of the QCD two-loop corrections to the form factors for the production of 
heavy quarks in $e^+ e^-$ collisions. 
In \cite{us1}, the vector current was considered (see also 
\cite{VEC}). Using the Laporta algorithm \cite{Laporta} for the
reduction of the dimensionally-regularized scalar integrals to the
set of master integrals and the differential equations technique
\cite{diffeq} for their calculation \cite{BMR1}, the ${\mathcal
O}(\alpha_S^2)$ corrections to the form factors for the vertex $\gamma Q
\bar{Q}$ were evaluated in terms of 1-dimensional harmonic polylogarithms 
\cite{hpls}.
In \cite{us2,us3}, the form factors for the axial vector, flavour singlet
axial vector and flavour non-singlet axial vector currents were calculated 
using the same technique (see also \cite{AX}). Particular attention was 
payed to the prescription for the $D$-dimensional extension of the 
$\gamma_5$ Dirac matrix. A pragmatic approach was used:
the diagrams not involving a closed triangular loop of fermions were 
evaluated performing the traces over the Dirac spinors with a naive 
anticommuting $\gamma_5$ \cite{us2}, while the anomalous diagrams were 
evaluated in \cite{us3} with the prescription proposed in \cite{HV,Larin}. 
In all cases the corresponding Ward identities were checked and found to 
be fulfilled. In particular, the use of the prescription of \cite{HV,Larin} 
in the evaluation of the anomalous diagrams in \cite{us3} breaks explicitly
the anomalous Ward identities. These have to be restored, by performing 
a finite renormalization. The constant terms for the finite renormalization 
were calculated and their equality to the ones for the massless case, given 
in the $\overline{\mathrm{MS}}$ scheme in \cite{Larin}, was verified. 
Moreover, in order to check Ward identities, the pseudo-scalar form factor 
for the corresponding diagrams and the truncated matrix element of the gluonic
operator $G \tilde{G}$ between the vacuum and an on-shell $Q \bar{Q}$ pair
state were calculated in \cite{us3}.
In \cite{us4}, finally, the evaluation of the scalar and pseudo-scalar form
factors was carried out in view of a completely differential description of the
decay of a neutral Higgs boson, which can couple both to scalar and
pseudo-scalar fermionic currents, into heavy quarks. 

The calculation of the form factors is only a part of the complete 
determination of the heavy quark production matrix element.
Nevertheless, it can also give very useful pieces of 
information on the precise determination of electroweak observables 
like, for instance, the forward-backward asymmetry $A_{FB}$ for heavy 
quarks at $e^+ e^-$ colliders \cite{Stefano}.
Moreover, heavy quark form factors can give hints and restrictions 
in the search of new physics. If
the mechanism of mass generation differs from the standard Higgs mechanism, 
deviations from the usual SM couplings of heavy particles  ($b$ or $t$ 
quarks) to photons and $Z$-bosons could be found. In \cite{us6}, the
NNLO QCD corrections to the anomalous magnetic moment and the weak 
axial-vector charge of $b$ and $t$ quarks were analyzed.
It was found, in particular, that the upper bound on the $b$ quark magnetic 
moment coming from LEP1 data is saturated by the corrections
due to two-loop perturbative QCD, which leave a limited room for new physics 
contributions to this quantity.
Finally, the form factors give the possibility to calculate the two-loop 
QCD corrections to the cross section of $e^+ e^- \to \gamma^{*},Z^{*}
\to t \bar{t}$ near the production threshold (c.f. \cite{HQthr} and refs.
therein). This physical observable is important, for instance, for the 
precise determination of the top quark mass that enters in the determination 
of the constraints given by the SM on the Higgs mass. Although QCD perturbative 
calculations do not allow an investigation of the cross section directly at 
threshold 
because of the presence of Coulomb divergences, we can get precise 
information in the region $\alpha_S \ll \beta \ll 1$, $\beta = \sqrt{1-4m^2/s}$ 
being the relative velocity of the quarks.
Moreover, QCD perturbative calculations are important to extract the 
matching coefficients in the NRQCD perturbative series in powers of 
$\alpha_S$ and $\beta$.

Here we present for brevity only the cross section to  
${\mathcal O}(\alpha_S^2)$ at the production threshold of the $Q \bar{Q}$ pair,
due to $\gamma$ and  $Z$ boson exchange. Omitting in the threshold 
expansion terms of ${\mathcal O}(\beta)$ (modulo the term $\sigma^{(0,Ve)}$ 
in Eq. (\ref{CS})) the $Q {\bar Q}$ contribution to the cross section 
is infrared finite by itself. Then we get, at NNLO,  the following 
analytic result:
\bea
\sigma  & = &  \sigma^{(0,Ve)} \Bigl\{ 1  +  \Delta^{(0,Ax)}  
+  C_{F} \frac{\alpha_{S}}{\pi} \Delta^{(1,Ve)}  \bigl( 1  
+  \Delta^{(0,Ax)} \bigr) \nn\\
& & \hspace*{1.71cm}
+ C_{F} \left( \frac{\alpha_{S}}{\pi} \right)^2  \bigl[ \Delta^{(2,Ve)}
\bigl( 1  +  \Delta^{(0,Ax)} \bigr) 
+  \Delta^{(2,Ax)}\bigr] \Bigr\} ,
\label{CS}
\eea
where the terms $\sigma^{(0,Ve)}$, $\Delta^{(1,Ve)}$, and $\Delta^{(2,Ve)}$ 
are the contributions of the photon exchange to the tree-level, one-loop and 
two-loop cross section respectively (their analytic expressions to 
${\mathcal O}(\alpha_{S}^{2}\beta^0)$ can be found in \cite{mel2}).
$\Delta^{(0,Ax)}$ and $\Delta^{(2,Ax)}$ are the contributions, at the same 
order in $\alpha_S$ and $\beta$, of the $Z$-boson exchange and the $Z-\gamma$ 
interference, normalized to $\sigma^{(0,Ve)}$:
\bea
\Delta^{(0,Ax)}  & = & 
    \frac{8 m^2 v_{Q}}{e_{Q} s_{W}^{2} c_{W}^{2} (4 m^2  -  m_{Z}^2)} 
    \Biggl[ - v_{e}  +  \frac{2 m^2 (v_{e}^{2}+a_{e}^{2}) 
    v_{Q}}{e_{Q} s_{W}^{2} c_{W}^{2} 
    (4 m^2 - m_{Z}^2)} \Biggr]  \nn\\
& &  
- \beta^2 \, \frac{4 m^2}{3 e_{Q} s_{W}^{2} c_{W}^{2} (4 m^2  -  m_{Z}^2)} 
    \Biggl\{  6 v_{e} v_{Q}
  - \frac{m^2}{(4 m^2  -  m_{Z}^2)} 
    \Biggl[ 24 v_{e} v_{Q}   \nn\\
& &  \hspace*{3cm}
  +  \frac{8 (v_{e}^{2}+a_{e}^{2})}{e_{Q} 
    s_{W}^{2} c_{W}^{2}} \Biggl( 3 v_{Q}^2 + a_{Q}^2 
    - \frac{12 m^2 v_{Q}^2}{(4 m^2 - m_{Z}^2)} \Biggr) \Biggr] \Biggr\} ,
\label{Z1} \\
\Delta^{(2,Ax)}  & = &  
   \frac{16 \zeta(2) m^4 a_{Q}^{2} ( v_{e}^{2}+a_{e}^{2} )}{e_{Q}^2 
   s_{W}^{4} c_{W}^{4} 
   (4 m^2 \! - \! m_{Z}^2)^2} \, C_F \, . 
\label{Z2} 
\eea
In Eqs. (\ref{Z1}), (\ref{Z2}) $m$ and $m_Z$ are the mass of the heavy quark
and of the $Z$-boson respectively, $s_W$ ($c_W$) the sine (cosine) of the weak 
mixing angle, $e_Q$ the charge of the heavy quark in units of the positron 
charge and $v_f = \frac{1}{2} ( I_{f}^{(3)} \! - 2 e_{f} s_{W}^2 )$, 
$a_f = \frac{1}{2} I_{f}^{(3)}$, with $f=e,Q$. The renormalization scale is 
taken equal to the heavy quark mass, $\mu=m$.
These expressions are obtained expanding the exact result for the virtual cross 
section in powers of $\beta$, retaining only terms of ${\mathcal O}(\beta^0)$. 
The diagrams that contribute to Eqs. (\ref{Z1}), (\ref{Z2}) are the ones 
of \cite{us1,us2}. It turns out, in fact, that at 
${\mathcal O}(\beta^0)$, the anomalous diagrams do not play any role.
The axial-vector contribution at ${\mathcal O}(\alpha_S^2)$, which can be 
obtained to all orders in $\beta$ from the results of \cite{us2}, is a new 
result. 
Here we give the leading term,
i.e., the term proportional to $a^2_Q$ in Eq. (\ref{Z2}) and in the 
${\mathcal O}(\beta^2)$ of Eq. (\ref{Z1}).
Notice that this term is of ${\mathcal O}(\beta^0)$, while the 
axial-vector contribution at ${\mathcal O}(\alpha_S)$ is of 
${\mathcal O}(\beta)$. The term
$\Delta^{(2,Ax)}$ is small compared to $\Delta^{(2,Ve)}$. Nevertheless, at a 
high luminosity linear collider with polarized $e^-$ and $e^+$ beams one may 
eventually be able to disentangle the vector and axial-vector induced 
contributions to the $t \bar t$ cross section. (For a calculation
of axial vector contributions 
in the context of Lippmann-Schwinger 
equations, 
see the 2nd reference of  \cite{AX}.)

In conclusion we have computed  the vector, axial-vector, scalar and 
pseudo-scalar vertices involving a pair of heavy quarks, $Q \bar{Q}$,
to ${\mathcal O}(\alpha_S^2)$, for arbitrary momentum transfer and mass of the 
heavy quarks. These form factors have a number of applications, including 
anomalous couplings, the $e^{+}e^{-} \to Q \bar Q$ cross section, the 
forward-backward asymmetry of heavy quarks, and the differential description
of Higgs boson decays into  $Q \bar Q$ pairs. Here we have briefly discussed 
the second order axial vector contributions to  the $Q {\bar Q}$ threshold
cross section, which is a new result. Although small these contributions 
may eventually be extracted from the $t \bar t$ threshold cross section 
measured at a high luminosity linear collider with polarized beams.

\section*{Acknowledgment}
This work was supported by the European Union under the contract 
HPRN-CT2002-00311 (EURIDICE) and by MCYT (Spain) under Grant FPA2004-00996, by
Generalitat Valenciana (Grants GRUPOS03/013 and GV05/015); by Deutsche 
Forschungsgemeinschaft (DFG), SFB/TR9, by DFG-Graduiertenkolleg RWTH Aachen,
by the Swiss National Science Foundation (SNF) under contract 200021-101874,
and by the USA DoE under the grant DE-FG03-91ER40662, Task J.


\end{document}